\documentclass[a4paper]{jpconf}
\usepackage{graphicx}
\def\lsi{\raise0.3ex\hbox{$<$\kern-0.75em\raise-1.1ex\hbox{$\sim$}}}
\def\gsi{\raise0.3ex\hbox{$>$\kern-0.75em\raise-1.1ex\hbox{$\sim$}}}
\newcommand{\lesim}{\mathop{\lsi}}

\def\be{\begin{equation}}
\def\ee{\end{equation}}
\def\ba{\begin{eqnarray}}
\def\ea{\end{eqnarray}}

\begin{document}

\title{Sterile neutrinos in cosmology and how to find them in the
lab
%}
\footnote{Invited talk at XXIII Int. Conf. on
Neutrino Physics and Astrophysics, May 25-31,
Christchurch, New Zealand.}}

\author{Mikhail Shaposhnikov}

\address{Institut de Th\'eorie des Ph\'enom\`enes Physiques, EPFL, 
CH-1015 Lausanne, Switzerland}

\ead{Mikhail.Shaposhnikov@epfl.ch}

\begin{abstract}
A number of observed phenomena in high energy physics and cosmology
lack their resolution within the Standard Model of particle physics.
These puzzles include neutrino oscillations, baryon asymmetry of the
universe  and existence of dark matter. We discuss the  suggestion
that all these problems can be solved by new physics which exists only
below the electroweak scale. The dedicated experiments that can
confirm or rule out this possibility are discussed.
\end{abstract}

%\section{Introduction}
%\label{sec:intro}
%
{\bf Introduction.} The  aim of this talk is to argue that the
existing high intensity protons beams, such as  NuMi beam at FNAL,
CNGS beam at CERN  and future accelerator facilities like J-PARC in
Japan, Project X at FNAL can be used to search for physics beyond the
SM in new dedicated experiments. A  {\em possible}  outcome of these
new experiments could be a discovery of  new neutrino states --
massive neutral leptons, new insight to the origin of neutrino masses,
fixing the pattern of neutrino mass hierarchy, and, eventually, 
discovery of  CP-violation in neutrino sector and revealing the origin
of baryon asymmetry of the universe and fixing its sign. The {\em 
guaranteed} outcome of these new experiments is the improving of the
constraints on the couplings of new particles by several orders of
magnitude.

The outline of the paper is as follows: first, we will discuss 
theoretical motivation for existence of relatively light singlet
leptons (they can be called singlet fermions, right-handed or sterile
neutrinos). It comes from the discovery of neutrino masses,  from
existence of dark matter (DM) and from baryon asymmetry of the
universe (BAU). Then we summarize the predictions of the properties of
singlet fermions  and  describe the strategy for the search for these
particles at existing and future accelerators. 

%\section{Neutrino masses}
%\label{sec:masses}	 
%
{\bf Neutrino masses.} Neutrinos have mass. A possible origin of this
mass is the existence of right-handed neutrinos $N_I$ with masses
$M_I$, $I=1,...,{\cal N}$. The most general renormalizable Lagrangian
incorporating the fields of the Standard Model (SM) and singlet
fermions has the form
\be
L=L_{\rm SM}+
\bar N_I i \partial_\mu \gamma^\mu N_I
  - F_{\alpha I} \,  \bar L_\alpha N_I \tilde \Phi
  - \frac{M_I}{2} \; \bar {N_I^c} N_I + h.c.,
  \label{lagr}
\ee
where $L_{\rm SM}$ is the Lagrangian of the SM,  $F_{\alpha I}$ are
the new Yukawa couplings, and $\Phi$ is the Higgs boson, $\tilde
\Phi_i = \epsilon_{ij}\Phi_j^*$. If the Dirac masses $M_D=F_{\alpha
I}v$ ($v= 174$ GeV is the vacuum expectation value of the Higgs field)
are much smaller than Majorana masses $M_I$, the type I see-saw 
formula holds $M_\nu = - M_D \frac{1}{M_I} [M_D]^T$ (for a review see
\cite{Strumia:2006db}). The number  of right-handed singlet fermions
must be at least two.  If there is only one of them, then two active
neutrinos are massless, which is at odds with the data on neutrino
masses and mixing. Already for ${\cal N}=2$ the Lagrangian
(\ref{lagr}) can describe the pattern of neutrino masses and mixings
observed experimentally. One of the most important parameters of
(\ref{lagr}) is the scale of the Majorana neutrino masses. However,
this parameter cannot be fixed by knowing $M_\nu$: multiply $M_D$ by
any number $x$ and $M_I$ by $x^2$ -- $M_\nu$ does not change.
Therefore, the choice of $M_I$ cannot be fixed by doing experiments
with active neutrinos only. 

{\em The GUT see-saw.} A popular choice for this scale is based on the
following logic.  {\em Assume} that Yukawa couplings of $N_I$ to the
Higgs and left-handed lepton doublets are similar to those in quark or
charged lepton sector (say, $F_{\alpha I} \sim F \sim 1$, as for the
top quark) and find $M_I$ from requirement that one gets correct
active neutrino masses: $M_I \simeq \frac{F^2 v^2}{m_{atm}}\simeq
6\times 10^{14}~{\rm GeV}$, where $m_{atm}\simeq 0.05$ eV is the
atmospheric neutrino mass difference. This scale happens to be close
to the scale of Grand Unification. There are theoretical challenges in
the GUT see-saw scenario. One of them  is related to the  hierarchy
problem: the mass  $M_I$ is much larger than electroweak (EW) scale.
Therefore, one should understand not only why $M_W \ll M_{Pl}$
($M_{Pl}=1.2\times10^{19}$ GeV is the Planck scale, $M_W$ is the mass
of the electroweak vector boson), but also why  $M_W \ll M_I$ and why
$M_I \ll M_{Pl}$. The smallness of the Higgs mass in comparison with
$M_I$ would require an extra fine-tuning \cite{Vissani:1997ys}.

{\em The EW see-saw} (for a review see \cite{Shaposhnikov:2007nf}).
{\em Assume}  that the Majorana masses of $N_I$  are smaller or of the
same order as the mass of the Higgs boson and find Yukawa couplings
from requirement that one gets the correct active neutrino masses: $F
\sim \frac{\sqrt{m_{atm} M_I}}{v} \sim 10^{-6}-10^{-13}$.  The EW
see-saw does not introduce any new energy scale besides the one
already present in the SM, and, therefore, contains no 
new hierarchy or fine tuning problem in comparison with the SM. This
allows a different approach to hierarchy problem, discussed in
\cite{Shaposhnikov:2007nj}. Though the stability of the Higgs mass
against radiative corrections gives a theoretical preference to the EW
see-saw, the low-energy neutrino experiments are indifferent to the
scale of $M_I$. Therefore, we add below two extra pieces of evidence
in favour of EW see-saw, coming from cosmology.

%\section{Dark matter}	
%\label{sec:DM} 
%
{\bf Dark matter.} About 23\% of the energy in the universe is
associated with non-baryonic DM. Amazingly, the theory (\ref{lagr})
gives a candidate for dark matter particle, provided one of the
singlet fermions is light enough (for a review see
\cite{Shaposhnikov:2007nj} and references therein). Indeed, if the
Yukawa couplings are small as in the EW see-saw, the lightest sterile 
neutrino $N_1$ can be practically stable and have a  lifetime which
may exceed greatly the age of the universe. 

\begin{figure}
\centerline{\includegraphics[width=8cm]{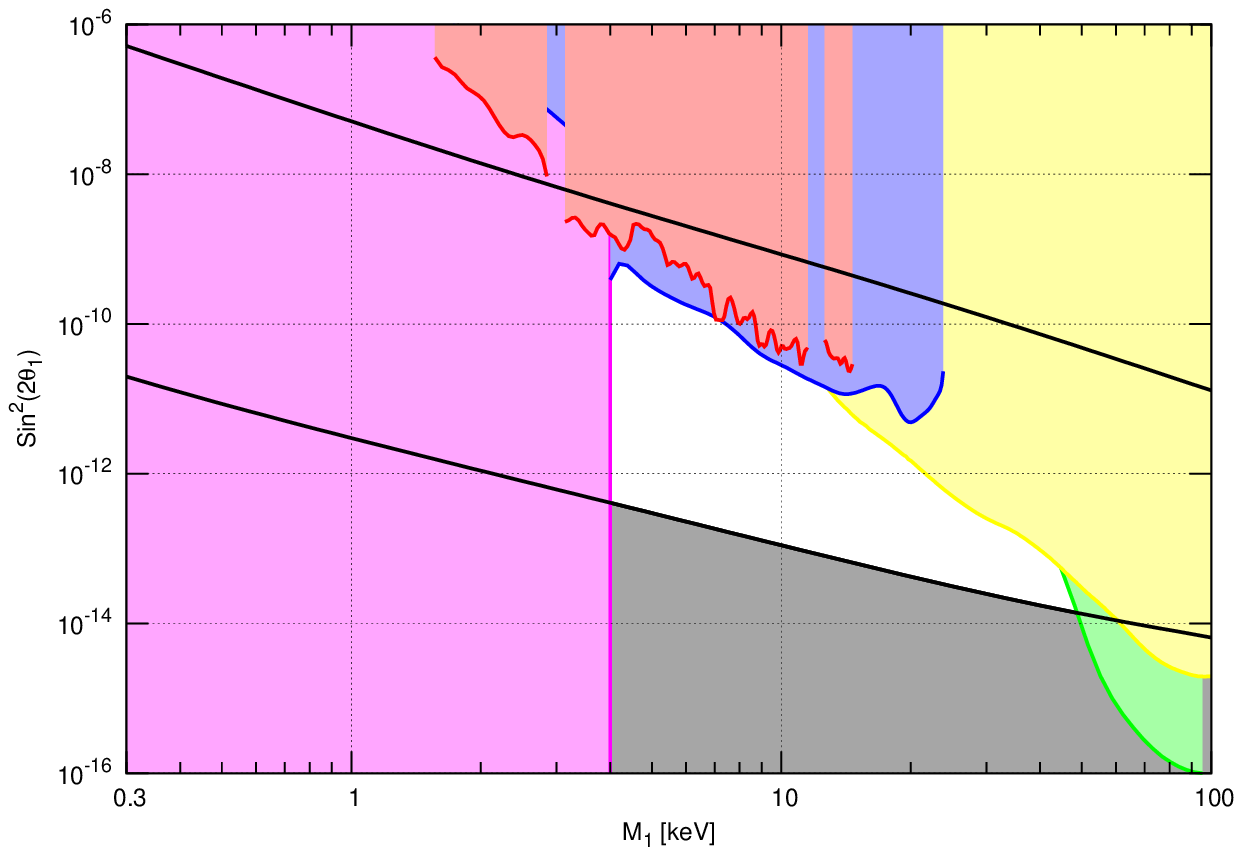}
\put(-52,105){\small$\Omega > \Omega_{DM}$}
\put(-75, 30){\small$\Omega < \Omega_{DM}$}
\put(-50, 78){\small$N_1 \to \nu\gamma$}
\put(-160, 80){\small Lyman-$\alpha$}
~~~
\includegraphics[width=8cm]{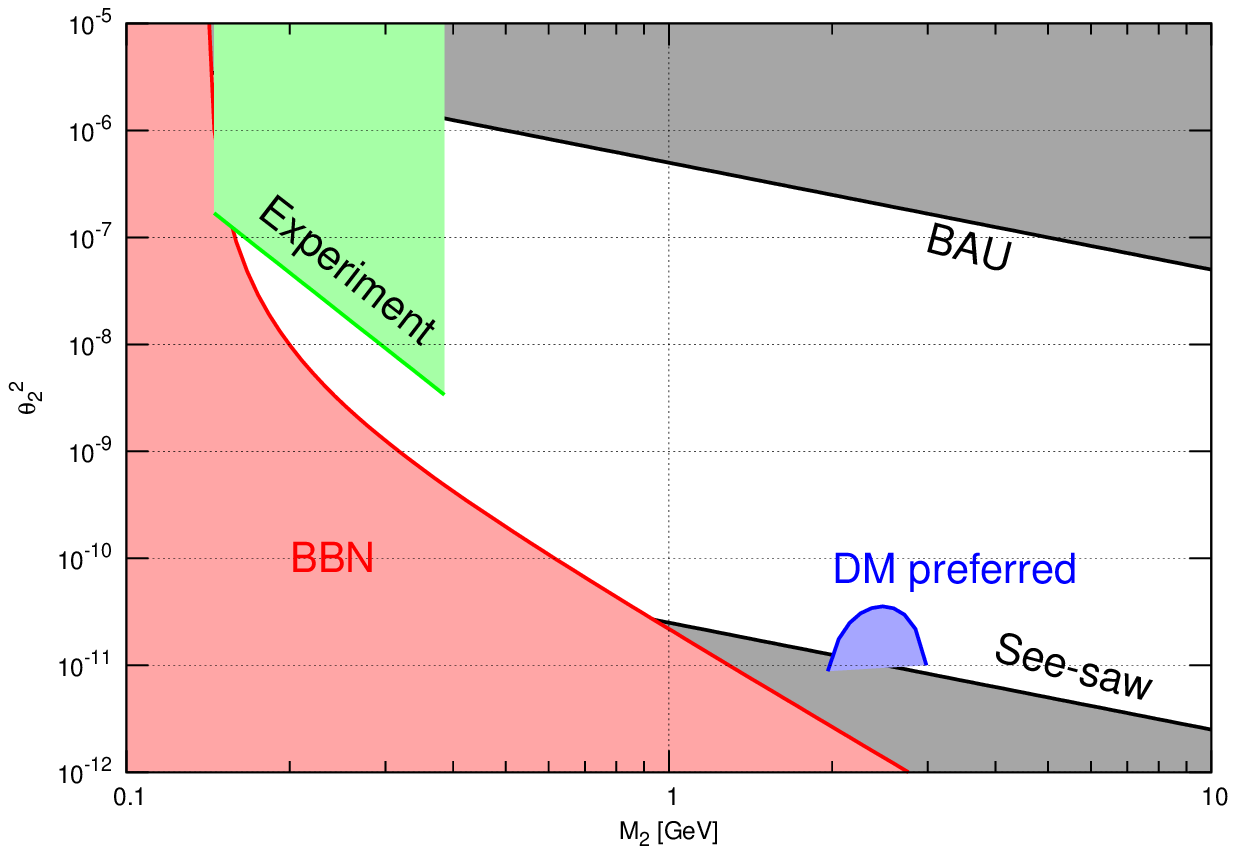}}
\caption{Left: constraints on the mixing angle $\theta_1 =
\frac{m_D}{M_I}$ of DM sterile neutrino. Right: constraints on the
mixing angle of BAU generating singlet fermions.} 
\label{fig:sterile}
\end{figure}

There are several constraints on sterile neutrino as a DM
candidate. They are shown in Fig.~\ref{fig:sterile} (left panel).
First, due to reactions $l\bar l \to \nu N_1,~~q\bar q \to \nu N_1$ 
etc, sterile neutrinos are created in the early universe. Their
abundance must correctly reproduce the measured density of DM.
Depending on other parameters of the Lagrangian (\ref{lagr}), the
admitted region lies between two black thick lines in 
Fig.~\ref{fig:sterile} \cite{Asaka:2006rw}. Second, the DM sterile
neutrino has a  sub-dominant radiative decay channel $N_1\rightarrow
\nu\gamma$, producing a narrow photon line which can be detected by
different X-ray satellites (for a review see \cite{Ruchayskiy:2007pq}
and references therein).  This line has not been seen. The right upper
corner in Fig.~\ref{fig:sterile} corresponds to the forbidden region,
coming from the analysis of a number of astronomical objects by
different X-ray instruments. Finally, a lower limit on the mass of DM
sterile neutrino comes from  structure formation. If $N_1$ is too
light it may have considerable free streaming length and erase
fluctuations on small scales. This can be checked by the study of 
Lyman-$\alpha$ forest spectra of distant quasars \cite{Seljak:2006qw}.
The region to the left of the vertical line corresponds to the
excluded region \cite{Asaka:2006rw}, which accounts for a non-trivial
velocity dispersion of DM particles.

An interesting feature of Fig. \ref{fig:sterile} is that the admitted
region is surrounded by different constraints in all directions,
telling that the hypothesis of sterile neutrino as a DM
candidate is experimentally testable. Moreover, the ${\cal O}(10)$ keV
scale for the mass of DM is singled out by these considerations. 

An important consequence of the cosmological and astrophysical
constraints on DM sterile neutrino is that $N_1$ does not contribute
significantly to the see-saw formula. Therefore, the number of singlet
fermions must be at least 3, to explain the DM and observed pattern of
neutrino masses and mixing angles. Since the number of fermion
families in the SM is 3, we take ${\cal N}=3$ in what follows, making
the particle content of the theory (we will call it the $\nu$MSM for
Neutrino Minimal Standard Model) similar in the left-handed and
right-handed sectors. 

Besides being a candidate for DM particle, sterile neutrinos may have
other interesting applications in astrophysics (for a review see 
\cite{Kusenko:2006zc}).

%\section{Baryon asymmetry}
%\label{sec:BAU} 	 
%
{\bf Baryon asymmetry.} Our universe is baryon asymmetric - it does
not contain antimatter in amounts comparable with matter. Quite
interestingly, the theory (\ref{lagr}) allows for generation of BAU
for a large choice of parameters of the model, and in particular for
wide range of masses of singlet fermions.

The case of GUT see-saw was discussed in talk by Y.~Nir at this
Conference. So, we elaborate on the case of EW see-saw only.
Remarkably, a pair of nearly degenerate light singlet fermions
$N_{2,3}$ also leads to baryogenesis, but due to another mechanism,
related to coherent oscillations of right-handed neutrinos  (for a
review see \cite{Shaposhnikov:2007nj} and references therein). The
light $N_I$ enter into thermal equilibrium very late due to the small
Yukawa couplings $F_{\alpha I}$. In particular, they may be out of
thermal equilibrium at all temperatures above $T_{EW}\sim 100$ GeV,
ensuring in this way one of the Sakharov conditions. The coherent
character of  oscillations leads to amplification of CP-violating
effects, to generation of lepton asymmetry and eventually to its
transfer to baryons because of non-perturbative EW effects.

In Fig. \ref{fig:sterile} (right panel) we present different
constraints on singlet fermion mixing angle versus their mass. Above
the lined marked ``BAU'' baryogenesis is not possible: here the
coupling of $N_{2,3}$ to active neutrinos is so large that they come
to thermal equilibrium above the EW temperature. Below the line marked
``See-saw'' the data on neutrino masses and mixings cannot be
explained. The region noted as ``BBN'' is disfavoured by the
considerations of Big Bang Nucleosynthesis - the decays of $N_{2,3}$
must not spoil the standard picture. A small region with the capture
``DM preferred'' in the domain of masses $2-3$ GeV  is quite peculiar:
here the generation of BAU  above the EW scale and production of DM
well below $T_{EW}$ is due to essentially the same mechanism, giving a
hint why the DM abundance is similar to that of baryonic matter.
Finally, the region marked ``Experiment'' shows the part of the
parameter space excluded by direct searches for singlet fermions. The
analysis of the published works of different collaborations reveals
that for the mass of the neutral lepton $M > 450$ MeV none of the past
or existing experiments enter into interesting for $\nu$MSM region
below the line ``BAU''. The NuTeV  upper limit on the mixing is at
most $10^{-7}$ in the region $M\simeq 2$ GeV \cite{Vaitaitis:1999wq},
whereas the NOMAD \cite{Astier:2001ck} and L3 LEP experiment
\cite{Achard:2001qw} give much weaker constraints.  The best
constraints in the small mass region, $M < 450$ MeV are coming from
the CERN PS191 experiment \cite{Bernardi:1985ny}, shown in 
Fig.~\ref{fig:sterile}. 

%\newpage
%\section{Summary of constraints on the parameters of the $\nu$MSM and
%its predictions} 
%\label{sec:predictions}
%
{\bf Summary of constraints on the parameters of the $\nu$MSM and its
predictions.} The first prediction is the absolute values of masses of
active neutrinos. One of the active neutrinos must be very light, $m_1
\lesim {\cal O}(10^{-6})$~eV. This fixes the masses of two other
active neutrinos: $m_2\simeq   9\cdot 10^{-3}$ eV, $m_3\simeq 5\cdot
10^{-2}$ eV for normal hierarchy or $m_{2,3}\simeq 5 \cdot 10^{-2}$ eV
for the inverted hierarchy. As a result, an effective Majorana mass
for neutrinoless double beta decay can be determined
\cite{Bezrukov:2005mx}. For normal (inverted) hierarchy the
constraints read:  $1.3~{\rm meV} <m_{\beta\beta}< 3.4~{\rm meV}$ 
($13~{\rm meV}< m_{\beta\beta}< 50~{\rm meV}$). A very conservative
bound  on the mass of DM sterile neutrino comes from analysis of
rotational curves of dwarf galaxies and reads $M_1 > 0.4$ keV
\cite{Boyarsky:2008ju} (it is weaker than the one coming from
Lyman-$\alpha$ discussed above). Direct experimental searches and BBN
require $M_{2,3} > 140~{\rm MeV}$ \cite{Gorbunov:2007ak}, whereas
baryogenesis due to sterile neutrino oscillations is possible if
$\Delta M=|M_2-M_3| < 800~ m_{atm}\left(M/{\rm GeV}\right)^2$
\cite{Asaka:2006rw}.

With quite a weak assumption about the initial conditions for the Big
Bang (no sterile neutrinos at the beginning (this assumption is
realized in the $\nu$MSM where the Higgs field plays the role of the
inflaton \cite{Bezrukov:2007ep}) the predictions and constraints can
be strengthened further. Namely the DM sterile neutrino mass should be
in the interval $4~\rm keV < M_1 < 50$ keV (the lowest bound is
related to Lyman-$\alpha$ observations), the DM sterile neutrino
mixing angle is predicted to be in the region $2\times 10^{-15} <
\theta^2_1 < 2\times 10^{-10}$. To produce the DM and BAU in correct
amounts, the mass of heavier neutral leptons should be in the region
$M_{2} \sim 2$ GeV, their level of degeneracy is constrained as
$\Delta M \lesim 10^{-4} m_{atm}$, and their mixing angle should be
$\theta^2_{2} \simeq 10^{-11}$. The CP asymmetry in $N_{2,3}$ decays
should be on the level of 1\% \cite{Asaka:2006rw}.

A direct experimental confirmation of the $\nu$MSM would be a
discovery of DM sterile neutrino and a pair of highly degenerated
neutral leptons. We will discuss below how these particles could be
searched for.

%\section{How to search for new leptons responsible for BAU
%\cite{Gorbunov:2007ak}} 
%\label{sec:lab}
%
{\bf The search for new leptons responsible for BAU
\cite{Gorbunov:2007ak}.} Let us consider a pair of heavier singlet
fermions, $N_2$ and $N_3$. Naturally, several distinct strategies can
be used for the experimental search of these particles.

The first one is related to their production ($\theta^2$ effect). The
singlet fermions  participate in all reactions the ordinary neutrinos
do with a probability suppressed roughly by a factor $\theta^2_{2}$.
Since they are massive, the kinematics of, say, two body decays $K^\pm
\rightarrow \mu^\pm N$, $K^\pm \rightarrow e^\pm N$ or three-body
decays  $K_{L,S}\rightarrow \pi^\pm + e^\mp + N_{2,3}$ changes  when
$N_{2,3}$ is replaced by an ordinary neutrino. Therefore, the study
of  {\em kinematics} of rare $K,~D$ and $B$ meson decays can constrain
the strength of the coupling of heavy leptons. This strategy has been
used in a number of experiments for the search of neutral leptons in
the past \cite{Yamazaki:1984sj,Daum:2000ac}, where the spectrum of
electrons or muons originating in decays $\pi$ and $K$ mesons has been
studied. The precise study of kinematics of rare meson decays is
possible in $\Phi$ (like KLOE), charm and B factories, or in
experiments with kaons where their initial 4-momentum is well known
(like NA48 or E787 experiments).

The second strategy is to use the proton beam dump ($\theta^4$
effect). As a first step the proton beam hitting the fixed target
creates $K,~ D$ or $B$ mesons which decay and produce $N_{2,3}$. The
second step is a search for decays of $N$ in a near detector, looking
for the processes ``nothing" $\rightarrow$ leptons and hadrons
\cite{Bernardi:1985ny,Vaitaitis:1999wq,Astier:2001ck}.   To this end
quite a number of already existing or planned neutrino facilities
(related, e.g. to CNGS, MiniBooNE, MINOS or J-PARC), complemented by a
near {\em dedicated} detector can be used. Finally, these two
strategies can be unified, so that the production and the decay occurs
inside the same detector \cite{Achard:2001qw}.

For the mass interval $M_I < M_K$ both strategies can be used.
Moreover, further constraints on the couplings of singlet fermions can
potentially be derived from the reanalysis of the {\em already
existing but never considered from this point of view} experimental
data of KLOE collaboration and of the E787 experiment. In addition,
the NA48/3 (P326) experiment at CERN and dedicated experiment at
MINER$\nu$A site, discussed by F. Vannucci at this Conference, would
allow to find or to exclude completely singlet fermions with the mass
below that of the kaon.

If $m_K < M_{2,3} < m_D$ the search for the missing energy signal,
potentially possible at beauty, charm and $\tau$ factories, is
unlikely to gain the necessary statistics and is very difficult if not
impossible at hadronic machines like LHC. So, the search for decays of
neutral fermions is the most effective opportunity.  The  dedicated
experiments on the basis of the proton beam NuMI or NuTeV at FNAL,
CNGS at CERN, or J-PARC  can touch a very interesting parameter range
for $M_I \lesim 1.8$ GeV. Experiments like NuSOnG (see talk by M.
Shaevitz an this Conference) and  HiResM$\nu$ \cite{petti} should be
able to explore a considerable part of the  cosmologically interesting
region for masses and mixing angles of singlet fermions.

Going above $D$-meson but still below $B$-meson thresholds is very
hard if not impossible with present or planned proton machines or
B-factories. To enter into cosmologically interesting parameter space
would require the increase of the present intensity of, say, CNGS beam
by two orders of magnitude or to producing and studying the kinematics
of more than $10^{10}$ B-mesons.  
\begin{figure}
\centerline{
\includegraphics[width=7.4cm]{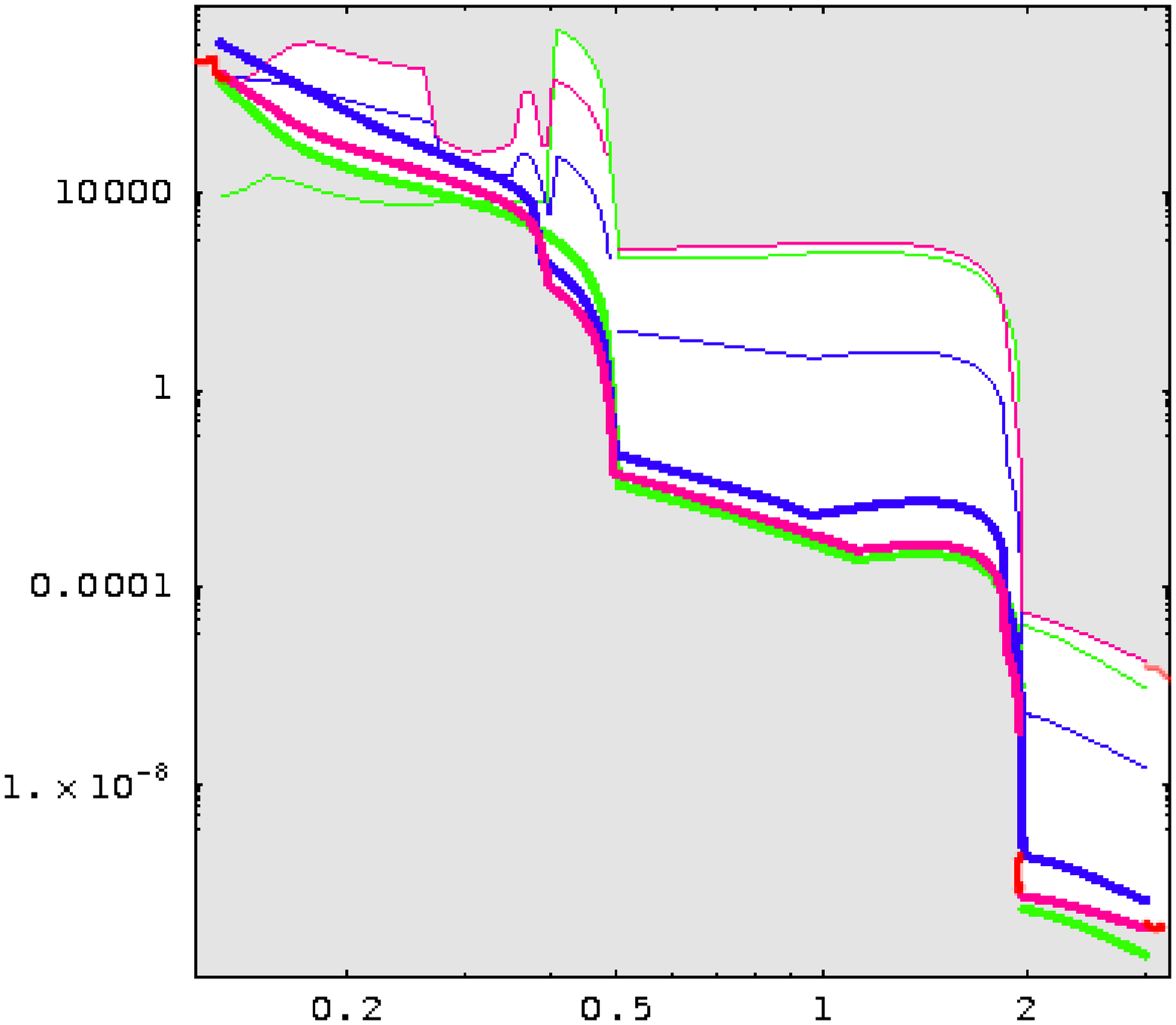}
\put(-80,165){\small  excluded by BAU}
\put(-140, 50){\small  excluded by $\nu$ masses}
~~~~~~~~
\raisebox{0.6cm}{
\includegraphics[width=7cm]{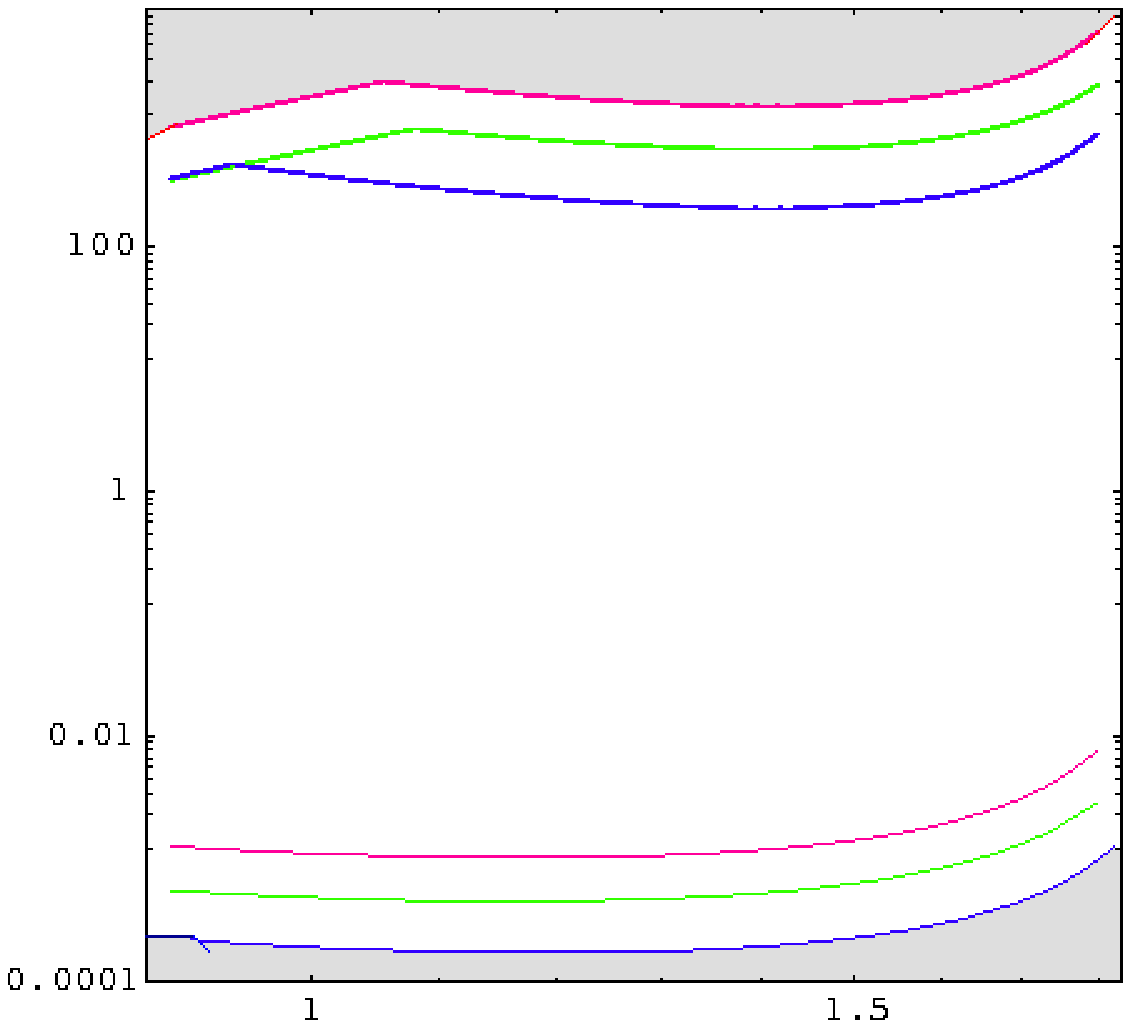}
\put(-95,144){\Huge $\uparrow$}
\put(-130,130){\small  excluded by $\nu$ masses}
\put(-130, 37){\small  excluded by BAU }
\put(-95,17){\Huge $\downarrow$}
}
}
\caption{Left panel: The number of singlet fermions decays expected in 5 m long
detector during one year with the use of J-PARC beam.
Right panel: The length of
detector in meters necessary to observe 10 decays of singlet fermions
per year if the proton beam of the Project X is used, as a function of
a mass. In the upper (lower - right panel) shaded area  baryogenesis
is not possible. In the lower (upper - right panel) shaded area neutrino
masses cannot be explained. Different curves correspond to the
different parameter choices in the $\nu$MSM.}
\label{fig:exp}
\end{figure}
In Fig. \ref{fig:exp} (left part)  we present the number of singlet
fermion decays  expected in 5 m long detector during one year with the
use of J-PARC beam  (the similar figures for CNGS,  NuMI and NuTeV can
be found in \cite{Gorbunov:2007ak}).  The right part of this figure
presents a length of detector necessary to observe 10 singlet fermion
decays per year in   X-Project beam-damp. In the upper (left panel)
and lower (right panel) shaded areas these particles cannot explain
BAU, and in the lower (left panel) and upper (right panel) shaded area
they cannot explain the observed neutrino masses and mixings.  The
Fig. \ref{fig:exp} shows that it is relatively easy to enter in the
region of the parameters interesting for cosmology, whereas it is very
challenging to explore all possible mixing angles of singlet fermions
below the charm threshold.

The couplings of $N_{2,3}$ are too small to see them at the LHC.
Inspite of this, the $\nu$MSM offers a specific prediction for the
search of new physics at the LHC experiments: nothing but the Higgs
in the mass interval   $M_H \in [129,189] ~~{\rm  GeV}$. This comes
about since in order to solve the SM problems (in particular, the one
related to inflation and to stability of the Higgs mass against
radiative corrections), the $\nu$MSM must be a valid field theory all
the way up to the Planck scale \cite{Shaposhnikov:2007nj}. Above the
upper limit the theory is not consistent due to Landau pole in the
scalar self-coupling (for a review see \cite{Hambye:1996wb}), whereas
below the lower limit the EW symmetry breaking vacuum is not stable
(for a review see \cite{Casas:1996aq}). 

%\section{Conclusions}  
%\label{sec:concl} 
% 
{\bf Conclusions.} New physics, responsible for neutrino masses and
mixings, for dark matter, and for baryon asymmetry of the universe may
hide itself {\em below} the EW scale. This possibility can be offered
by the  the $\nu$MSM - a minimal model, explaining simultaneously {\em
all well-established observational} drawbacks of the SM. 

This new physics (a pair of new neutral leptons, creating the baryon
asymmetry of the universe) can be searched for in dedicated
experiments with the use of existing intensive proton beams at CERN,
FNAL and planned neutrino facilities in Japan (J-PARC). An indirect
evidence in favour of this proposal will be given by LHC, if it 
discovers the Higgs  boson within  the mass  interval discussed above
and nothing else.  Moreover,  the $\nu$MSM gives a hint on how and
where to search for new physics in this case. It tells, in particular,
that in order to uncover new phenomena in particle physics one should
go towards high intensity proton beams or very high intensity charm or
B-factories, rather than towards high energy electron-positron
accelerators. 

To search for DM sterile neutrino in the universe one needs an X-ray
spectrometer in Space  with good energy resolution $\delta E/E \sim
10^{-3}-10^{-4}$ getting signals from our Galaxy and its dwarf
satellites \cite{Boyarsky:2006fg}. The laboratory search for this
particle would require an extremely challenging detailed analysis of
kinematics of $\beta$-decays of different isotopes
\cite{Bezrukov:2006cy}.

{\bf Acknowledgments.} This work was supported in part by Swiss
National Science Foundation. It is a pleasure to thank T. Asaka, F.
Bezrukov, S. Blanchet,  A. Boyarsky, D. Gorbunov, A. Kusenko, M.
Laine, A. Neronov, O. Ruchayskiy,  and I. Tkachev for collaboration on
the topics described in this talk.

\section*{References}

\vspace{0.5cm}

\end{document}